\documentclass[a4paper,aps,pra,reprint,twocolumn,floatfix,superscriptaddress,notitlepage,usenames,dvipsnames,svgnames,table,footinbib,longbibliography]{quantumarticle}
\pdfoutput=1
\usepackage[utf8]{inputenc}
\usepackage[T1]{fontenc}
\usepackage{graphicx}
\usepackage{grffile}
\usepackage{wrapfig}
\usepackage{rotating}
\usepackage[normalem]{ulem}
\usepackage{amsmath}
\usepackage{textcomp}
\usepackage{amssymb}
\usepackage{capt-of}

\usepackage{parskip}
\usepackage{mathrsfs}
\usepackage[margin= 0.75in]{geometry}
\usepackage[braket, qm]{qcircuit}

\usepackage[english]{babel}
\usepackage{letltxmacro}
\LetLtxMacro{\ORIGselectlanguage}{\selectlanguage}
\makeatletter
\DeclareRobustCommand{\selectlanguage}[1]{%
  \@ifundefined{alias@\string#1}
    {\ORIGselectlanguage{#1}}
    {\begingroup\edef\x{\endgroup
      \noexpand\ORIGselectlanguage{\@nameuse{alias@#1}}}\x}%
}
\newcommand{\definelanguagealias}[2]{%
  \@namedef{alias@#1}{#2}%
}
\makeatother
\definelanguagealias{en}{english}

\usepackage{bbm}
\usepackage{etoolbox}
\usepackage{tcolorbox} % if you want to make a fancy box around some equations
\usepackage{fleqn} % automatically adds () around equation refs.
\usepackage{tikz}
\usepackage{amsthm}
\usepackage{amssymb}
\usetikzlibrary{arrows.meta}
\usepackage{mathtools}

\newcommand{\prlsection}[1]{{\em {#1}.---~}}

\usepackage{bm}
\usepackage{dsfont}
\usepackage[font=small,labelfont=bf,justification=justified,format=plain]{caption}
\usepackage{subcaption}

\usepackage{thmtools}
\usepackage{thm-restate}
\definecolor{blue-violet}{rgb}{0.54, 0.17, 0.89}
\usepackage{hyperref}
\hypersetup{
 pdfauthor={The quantum coherent people},
 pdftitle={Open OTOC},
 pdflang={English},colorlinks=true,linkcolor=RubineRed,citecolor=blue-violet,urlcolor=Cerulean} 
\usepackage[capitalise]{cleveref}

\begin{document}
\title{Quantum Scrambling of Observable Algebras}

\author{Paolo Zanardi}
\email [e-mail: ]{zanardi@usc.edu}

\affiliation{Department of Physics and Astronomy, and Center for Quantum Information Science and Technology, University of Southern California, Los Angeles, California 90089-0484, USA}

% \date{\today}

\begin{abstract}
In this paper we describe an algebraic/geometrical approach to quantum scrambling. Generalized quantum subsystems are described by an hermitian-closed unital subalgebra $\cal A$ of operators  evolving through a unitary channel. Qualitatively, quantum scrambling is defined by how the associated physical degrees of freedom get mixed up with others by the dynamics.
Quantitatively, this is accomplished  by introducing a  measure, the  geometric algebra anti-correlator (GAAC),  of the self-orthogonalization of  the commutant of $\cal A$  induced by the dynamics. This approach extends and unifies averaged bipartite OTOC,  operator entanglement,  coherence generating power and Loschmidt echo. Each of these concepts is indeed recovered by a special choice of $\cal A$. 
We compute typical values of GAAC for random unitaries, we prove upper bounds and characterize their saturation. For generic energy  spectrum we find explicit expressions for the infinite-time average of the GAAC  which encode the relation between  $\cal A$ and the full system of Hamiltonian eigenstates. Finally, a notion of ${\cal A}$-chaoticity is suggested.
\end{abstract}
\maketitle
%\begin{widetext}
\prlsection{Introduction}
Quantum  dynamics can quickly spread information, which was initially  encoded in some physical degrees of freedom, into a larger set of degrees of freedom, in this way quantum information gets  \emph{delocalized} and highly non-local correlations can be built.  
This so-called \emph{quantum scrambling}, has over the last few years attracted  a growing amount  of attention in the context of quantum chaos  and also quantum computing. The Out of Time Order Correlation functions (OTOCs) are among the most popular tools to analyze scrambling from a quantitative point view \cite{larkin1969quasiclassical,kitaev_simple_2015,MaldacenaChaos2016,PhysRevLett.115.131603,PolchinskiSYK2016,MezeiChaos2017,Roberts2017Chaos}.

The goal of this paper is to lay down a novel  formalism for quantum scrambling. Roughly speaking,  we will characterize scrambling by  how much a \emph{whole} set of distinguished  degrees of freedom gets far from itself by unitary evolution.
 The underlying philosophy of this paper is an extension of the observable-algebra approach to quantum subsystems originally advocated in \cite{zanardi-virtual-2001,zanardi-tps-2004} (see also recent developments in \cite{kabernik-coarse-2018,kabernik-reduction-2020}). As such the strategy  can be applied to situations in which there is  no an \emph{a priori} locality structure which  gives a natural way of defining subsystems e.g., see \cite{carroll-mereology-2021}. % in the context of quantum gravity.
 
 We will show that specific instances of our construction  allow one to recover apparently different concepts including operator entanglement \cite{zanardi2001entanglement,Prosen_operator_2007},  averaged bipartite OTOCs \cite{yan2020information,styliaris_information_2020}, coherence generating power \cite{zanardiCoherencegeneratingPowerQuantum2017, zanardiMeasuresCoherencegeneratingPower2017, styliarisCoherencegeneratingPowerQuantum2018}    and Loschmidt echo \cite{jalabert2001environment,goussev2012loschmidt}. This  conceptual unification  provides one of the main motivations for this work.
Another one is to design candidate tools for unveiling novel facets of quantum chaos. 

For the sake of clarity, the main technical results of this paper are organized in ``Propositions''
whose proofs are  in \footnote{\label{footnote:SM} See supplemental material}.

\prlsection{Preliminaries}
\label{sec:Preliminaries}
In this section we introduce the main formal  ingredients utilized in this paper and set the notation.
Let ${\cal H}=\mathrm{span}\{|m\rangle\}_{m=1}^d$ be a $d$-dimensional Hilbert space and $L({\cal H}) $ its the full operator algebra (see  \footnote{
$L({\cal H}) $  has  a Hilbert space structure via  the Hilbert-Schmidt scalar product: $\langle X,\,Y\rangle:=\mathrm{Tr}\left(X^\dagger Y\right)$ and norm
$\|X\|_2^2:=\langle X,\, X\rangle.$
This equips the space of superoperators i.e., $L(L({\cal H}))$ with the scalar product $\langle {\cal T},\,{\cal F}\rangle:=\mathrm{Tr}_{HS}\left({\cal T}^\dagger {\cal F} \right)=
\sum_{l,m} \langle m| {\cal T}^\dagger {\cal F}(|m\rangle\langle l|)|l\rangle,$ and the norm $\|  {\cal T}\|_{HS}^2= \langle {\cal T},\,{\cal T}\rangle=\sum_{l,m} \|{\cal T}(|m\rangle\langle l|)\|_2^2.$
If ${\cal T}(X)=\sum_i A_i XA_i^\dagger,$ then $\|{\cal T}\|_{HS}^2=\sum_{i,j} |\mathrm{Tr}(A_i^\dagger A_j)|^2$} for further notation).
In the following by the notation ${\mathbf{C}}\{X\}$ we will denotes the vector space spanned by the $X'$s.

The key formal ingredients of this investigation are \emph{hermitian-closed  unital  subalgebras} ${\cal A}\subset L({\cal H})$ and their commutants ${\cal A}^\prime:=\{X\in L({\cal H})\,/\, [X,\,Y]=0,\,\forall Y\in {\cal A}\}.$
The intersection ${\cal A}\cap {\cal A}^\prime=:{\cal Z}({\cal A})$ is the \emph{center} of the algebra ${\cal A}.$ 
The fundamental structure theorem of these objects states that the Hilbert
space breaks into a direct sum of $d_Z:=\mathrm{dim} \,{\cal Z}({\cal A})$ orthogonal blocks  and each of them has a tensor product bi-partite structure:
${\cal H}=\oplus_J {\cal H}_J,\, {\cal H}_J,\cong \mathbf{C}^{n_J}\otimes  \mathbf{C}^{d_J}.$ Moreover, 
\begin{align}\label{eq:Alg-central-decomposition}
{\cal A}\cong \oplus_J \openone_{n_J}\otimes L( \mathbf{C}^{d_J}),\qquad{\cal A}^\prime\cong \oplus_J L( \mathbf{C}^{n_J})\otimes \openone_{d_J}.
\end{align}
 Whence, $d=\sum_Jn_Jd_J,$ $\mathrm{dim}\,{\cal A}=\sum_Jd_J^2=:d({\cal A})$ and $\mathrm{dim}\,{\cal A}^\prime=\sum_Jn_J^2=:d({\cal A}^\prime).$  
 Also, $ {\cal Z}({\cal A})=\mathbf{C}\{\Pi_J:=\openone_{n_J}\otimes \openone_{d_J}\},$ namely the center of $\cal A$ is spanned by the projections over the ${\cal H}_J$ blocks.

Associated to any algebra $\cal A$ we have an orthogonal (super) projection CP-map: $\mathbb{P}_{\cal A}^\dagger=\mathbb{P}_{\cal A},\, \mathbb{P}_{\cal A}^2=\mathbb{P}_{\cal A}$ and
$\mathrm{Im}\,\mathbb{P}_{\cal A}={\cal A}.$ Such maps can be written in the Kraus form  $\mathbb{P}_{\cal A}(X)=\sum_{\alpha=1}^{d({\cal A}^\prime)}e_\alpha X e_\alpha^\dagger,$ where
the $e_\alpha$ are a suitable {\em{orthogonal }} basis  of ${\cal A}^\prime.$
Notice that $\mathrm{Tr}_{HS} \mathbb{P}_{\cal A}=\sum_{\alpha=1}^{d({\cal A}^\prime)}|\mathrm{Tr}\, e_\alpha|^2=d({\cal A}).$
In terms of the decomposition (\ref{eq:Alg-central-decomposition}) one has 
$\mathbb{P}_{{\cal A}}(X) =\sum_J \frac{\openone_{n_J}}{n_J} \otimes \mathrm{tr}_{n_J}(X),$ and 
$\mathbb{P}_{{\cal A}^\prime}(X) =\sum_J \mathrm{tr}_{d_J}(X)\otimes \frac{\openone_{d_J}}{d_J}.$

These structural results provide the mathematical underpinnings of the theory of decoherence-free subspaces \cite{zanardi-noiseless-1997,lidar-dfs-1998},
noiseless subsystems \cite{KLV-2000,zanardi-stabilizing-2000} and in general to all quantum-information stabilizing techniques \cite{zanardi-stabilizing-2000}.
From the physical point of view two special cases are worth emphasizing: 

\emph{Factors}: ${\cal Z}({\cal A})=\mathbf{C}\openone,$ in this case ${\cal H}\cong \mathbf{C}^{n_1}\otimes\mathbf{C}^{d_1}$
namely the algebra $\cal A$ endows $\cal H$ with a bipartition into \emph{virtual} subsystems \cite{zanardi-virtual-2001,zanardi-tps-2004}. 
The case in which ${\cal H}={\cal H}_A\otimes {\cal H}_B$ with ${\cal A}=L({\cal H}_A)\otimes\openone_B$  and ${\cal A}^\prime=\openone_A\otimes L({\cal H}_B)$ clearly falls in this category.

\emph{Super-Selection}: ${\cal A}^\prime\subset {\cal A}$ this is when the commutant is an Abelian algebra. This implies $n_J=1,\,(\forall J)$
and therefore  the Hilbert space breaks into $d_J$-dimensional {\em super-selection} sectors i.e., ${\cal H}\cong \oplus_{J=1}^{d({\cal A}^\prime)} {\cal H}_J,\,{\cal H}_J\cong \mathbf{C}^{d_J}$
and ${\cal A}\cong \oplus_{J=1}^{d({\cal A}^\prime)} L(\mathbf{C}^{d_J}).$ 
%In this context the projections $\Pi_J$ are referred to as the \emph{super-selection charges.}
If $\cal A$ is a maximal abelian subalgebra  one has  ${\cal A}={\cal A}^\prime$ and $n_J=d_J=1,\,(\forall J).$
%2) ${\cal A}={\cal A}^\prime\cong \mathbf{C}^d$ this is when ${\cal A}$ is maximal abelian subalgebra of $L({\cal H}).$
%In this case $n_J=d_J=1$ for $J=1,\ldots,d.$ 
This is  the case  that is relevant to the study of quantum coherence \cite{streltsov_colloquium_2017} and its dynamical generation  \cite{zanardiCoherencegeneratingPowerQuantum2017, zanardiQuantumCoherenceGenerating2018}. 

When the $d_Z$-dimensional (integer-valued) vectors ${\bf{d}}:=(d_J)_J$, and ${\bf{n}}:=(n_J)_J$ are proportional to each other
i.e., ${\bf{d}}=\lambda {\bf{n}}$ one has that $d^2= d({\cal A})d({\cal A}^\prime).$ If this is the case we shall say that the pair $({\cal A},\,{\cal A}^\prime)$ is \emph{collinear}.
Note that both factors and  maximal abelian subalgebras are of this type.

%

%%%%%%%%%%%%%%%%%%%%%%% GENERAL RESULTS %%%%%%%%%%%%%%%%%%%%
\prlsection{General  results}
\label{sec:General-results}
%\begin{restatable}{prop}{U-invariance}
%\label{th:U-invariance}
%Let be ${\cal A}\subset L({\cal H})$ a $*$-closed subalgebra, $\mathbb{P}_{\cal A}$ the (super) projection onto it and ${\cal U}$ a unitary channel ($X\mapsto UXU^\dagger,\,(U\in U({\cal H})$)
%The following properties are equivalent
% ) ${\cal U}({\cal{A}}):=\{{\cal U}(X)\,/\, X\in{\cal A}\}={\cal A},$ ii) $[{\cal U},\,\mathbb{P}_{\cal A}]=0,$ iii) ${\cal U}\mathbb{P}_{\cal A}=\mathbb{P}_{\cal A}{\cal U}\mathbb{P}_{\cal A},$ iv) all the above with ${\cal A}\leftrightarrow{\cal A}^\prime.$
%\end{restatable}
We are now in the position to define the central mathematical object of this paper: the
 \emph{geometric algebra anti-correlator} (GAAC) by
\begin{align}
\label{eq:GAAC}
G_{\cal A}(U):=1 -\frac{\langle \mathbb{P}_{{\cal A}^\prime}, \,\mathbb{P}_{{\cal U}({\cal A}^\prime)} \rangle}{\|\mathbb{P}_{{\cal A}^\prime}\|_{HS}^2}.
\end{align}
The geometrical meaning of GAAC should be evident from Eq. (\ref{eq:GAAC}):  the larger $G_{\cal A}(U)$ the smaller is the intersection between ${\cal A}^\prime$ and its unitarily evolved image ${\cal U}({\cal A}^\prime):=\{ {\cal U}(X)\,/\, X\in {\cal A}^\prime\}$ \footnote{This can be seen from the fact that given two projectors  $P$, and $Q$ of rank $d$ one has:
$\mathrm{dim}(V_P\cap V_Q)\le \mathrm{tr}(PQ)\le d.$ Where $V_{P/Q}$ are the images of $P/Q.$
The lower (upper) bound is achieved when $P$ and $Q$ commute (coincide). }

\emph{Remark.--} In the RHS of Eq.~(\ref{eq:GAAC}) we use the ${\cal A}^\prime$ (and not  $\cal A$) as the dynamics $\cal U$ is in the Heisenberg picture. Symmetries mapped out of ${\cal A}^\prime$ by ${\cal U}$ is equivalent to states mapped out of $\cal A$ by ${\cal U}^\dagger.$
This choice is somewhat arbitrary (See Prop.~1).

%\footnote{A few remarks: 1) $\|\mathbb{P}_{{\cal A}^\prime}\|_{HS}^2=\mathrm{Tr_{HS}\mathbb{P}_{{\cal A}^\prime}}= \mathrm{dim} \,{\cal A}^\prime=:d({\cal A}^\prime).$  is the dimension of the commutant ${\cal A}^\prime.$ 2) $\mathbb{P}_{{\cal U}({\cal A}^\prime)}={\cal U} \mathbb{P}_{{\cal A}^\prime}{\cal U}^\dagger.$ 3) $\langle \mathbb{P}_{{\cal A}^\prime}, \, \mathbb{P}_{{\cal U}({\cal A}^\prime)}\rangle=\| \mathbb{P}_{{\cal A}^\prime}{\cal U} \mathbb{P}_{{\cal A}^\prime}\|_{HS}^2,$ has  range is between $1$ and $\|\mathbb{P}_{{\cal A}^\prime}\|_{HS}^2.$ Whereby $0\le G_{\cal A}(U)\le 1- 1/d({\cal A}^\prime).$ }. 

%We shall see more from this perspective later.
%
Algebraically, (\ref{eq:GAAC}) measures how much the symmetries of the generalized quantum subsystem associated to $\cal A$ are dynamically broken by the channel ${\cal U}.$ 
Let us now start by  further unveiling the geometrical nature of GAACs. 
First notice that, using the algebra super-projections, one can define a distance between two algebras $\cal A$ and $\cal B$: $D({\cal A},\,{\cal B}):=\|\mathbb{P}_{\cal A}- \mathbb{P}_{\cal B}\|_{HS}.$
This metric structure allows one to draw a quite simple geometrical  picture of algebra scrambling.
%From Eq. (\ref{eq:GAAC} it easily  follows  
\begin{restatable}{prop}{prop-0}\label{th:Algebra-invariance}
i) The GAAC  is the  (squared and normalized) distance between the algebra ${\cal A}^\prime$ and its image ${\cal U}({\cal A}^\prime).$
 \label{th:GAAC-is-distance}
\begin{align}\label{eq:GAAC-is-distance}
G_{\cal A}(U)=
\frac{1}{2}\frac{D^2\left({\cal A}^\prime,\, {\cal U}({\cal A}^\prime)\right)}{d({\cal A}^\prime)}
%=\frac{\|{\cal U}\mathbb{P}_{{\cal A}^\prime}-\mathbb{P}_{{\cal A}}^\prime}{\cal U}\mathbb{P}_{{\cal A}}^\prime} \|_{HS}^2}{\|\mathbb{P}_{{\cal A}^\prime}\|_{HS}^2}
\end{align} 
%where $D({\cal A},\,{\cal B}):= \|   \mathbb{P}_{{\cal A}} -\mathbb{P}_{\cal B} \|_{HS}^2$ is a distance function between the algebras ${\cal A}$ and ${\cal B}.$
ii) $G_{\cal A}(U)=0 \Leftrightarrow {\cal U}({\cal A}^\prime)={\cal A}^\prime  \Leftrightarrow   {\cal U}({\cal A})={\cal A}.$ In words: the GAAC Eq. (\ref{eq:GAAC}) vanishes if and only if both algebras  $\cal A$ and ${\cal A}^\prime$ are invariant under ${\cal U}$ i.e., there is no algebra scrambling.
\end{restatable}

%From this perspective the algebra scrambling quantified by the GAAC (\ref{eq:GAAC}) is very intuitive, i
%This proposition unveils the fundamental geometrical meaning of GAAC: it  is just how far the channel $\cal{U}$ sends ${\cal A}^\prime$ from itself normalized by  the commutant dimension $d({\cal A}^\prime).$
The definition of GAAC given by Eq. (\ref{eq:GAAC})  has the drawback of relying of superoperator projections and therefore may seem somewhat abstract and removed from practical calculations.
Hence, before moving on to physical examples and applications of our formalism, we would like to re-express the GAAC at the more familiar operator level.
 \begin{restatable}{prop}{Omega-picture}
 \label{th:Omega-picture}
i)  One can find an orthogonal basis of $\cal A$ $\{e_\alpha\}_{\alpha=1}^{d({\cal A})}$ and an orthonormal basis of ${\cal A}^\prime$ $\{f_\gamma\}_{\gamma=1}^{d({\cal A}^\prime)}$
 such that 
 \begin{align}\label{eq:Omega-picture}
1-G_{\cal A}(U)= \frac{\langle \Omega_{\cal A},\,{\cal U}^{\otimes\,2}(\Omega_{\cal A})\rangle}{\|\Omega_{\cal A}\|_2^2}
= \frac{\langle\tilde{\Omega}_{{\cal A}},\,{\cal U}^{\otimes\,2}(\tilde{\Omega}_{{\cal A}})\rangle}{\|\tilde{\Omega}_{{\cal A}}\|_2^2}
 \end{align} 
 where $\Omega_{\cal A}:=\sum_{\alpha=1}^{d({\cal A})} e_\alpha\otimes e^\dagger_\alpha,$ and 
$\tilde{\Omega}_{{\cal A}}=\sum_{\gamma=1}^{d({\cal A}^\prime)} f_\gamma\otimes f^\dagger_\gamma.$ Also, $\tilde{\Omega}_{{\cal A}}= S \Omega_{\cal A},$ where $S$ is the swap  
on     ${\cal H}^{\otimes\,2},$
 and $\|\Omega_{{\cal A}}\|_2^2=\|\tilde{\Omega}_{{\cal A}}\|_2^2=d({\cal A}^\prime).$
 
 ii) If $({\cal A},\,{\cal A}^\prime)$ is {collinear} then $G_{\cal A}(U)=G_{{\cal A}^\prime}(U),(\forall U)$.
\end{restatable}
%\mathbb{P}_{{\cal A}^\prime} =\mathbb{P}_{{\cal U}({\cal A}^\prime)}.$ 
%\end{widetext}
In the above proposition, all the  (Hilbert-Schmidt)  scalar products and norms are ordinary operators ones. Moreover, the $\Omega$'s operator
can be expressed in the same way if the bases $e_\alpha$'s and $f_\alpha$'s are replaced by unitarily equivalent ones. The connection between Eqs (\ref{eq:GAAC}) and (\ref{eq:Omega-picture})
is given by 
\begin{align}\label{eq:Omega-projections}
\mathbb{P}_{{\cal A}^\prime}(X)=\mathrm{Tr}_1\left(S \Omega_{\cal A} (X\otimes\openone)\right)=\mathrm{Tr}_1\left(\tilde{ \Omega}_{\cal A} (X\otimes\openone)\right).
\end{align}
Interestingly, the no-scrambling condition $G_{\cal A}(U)=0$ using Prop. \ref{th:Omega-picture} can be expressed by the  operator fixed-point equations
${\cal U}^{\otimes\,2}(\Omega_{\cal A})= \Omega_{\cal A}.$ %(same with $\Omega_{{\cal A}^\prime}.$) 
The (unsurprising) price to pay is that now the Hilbert space is doubled.
Another advantage of the formulation (\ref{eq:Omega-picture}) is that it makes clear that the GAAC can be computed in terms of  $2$-point correlation functions.
In fact, from  Eq. (\ref{eq:Omega-picture}) one finds (see appendix)
\begin{align}\label{2-point-GAAC}
1- G_{\cal A}(U)=\frac{1}{d({\cal A}^\prime)} \sum_{\alpha,\beta=1}^{d({\cal A})} |\langle e_\alpha,\,{\cal U}(e_\beta)\rangle |^2,
\end{align}
(a similar expression hold for the $f_\alpha$'s). This expression suggests how one could measure the GAAC by resorting to process tomography for $\cal U$.
Notice also that operational protocols to measure the GAAC were already discussed, for the cases {\bf{1)}} and {\bf{2)}} here below, in \cite{styliaris_information_2020} and \cite{zanardiCoherencegeneratingPowerQuantum2017} respectively.

\prlsection{Physical Cases}%%%%%%%%%%%%%%%%%%%%%%%%%%%%%%%%%%%%%%%%%%%%%%%%%%%%%%%%%%%%%%%%%% TWO PHYSICAL CASES %%%%%%%%%%%%
\label{sec=Physical-cases}
 To concretely illustrate the formalism let us now consider several  physically motivated  examples in which the GAAC can be fully computed analytically. 

The first two   examples show how the GAAC formalism   allows one to understand two ostensibly unrelated physical problems, operator entanglement \cite{zanardi2001entanglement} %[which coincides with the averaged bipartite OTOC \cite{yan_information_2019,styliaris_information_2020}] 
and  coherence generating power (CGP) \cite{zanardiCoherencegeneratingPowerQuantum2017, zanardi-CGP-measures-2007}, from a single vantage point.   
The first (second) concept is obtained when $\cal A$ is a factor (maximal abelian).
This means that one can also think of the GAAC either as an extension of operator entanglement to algebras that are not factors, or as an extension of coherence generating power to algebras that are not maximal abelian subalgebras.

The third and fourth examples  are "dual"  to each other and  show that, in general, $G_{\cal A}(U)\neq G_{{\cal A}^\prime}(U).$
Finally, the fifth illustrates  in which sense even the concept of \emph{Loschmidt echo}, a valuable tool in the study of quantum chaos  \cite{peres1984stability,PhysRevLett.86.2490,Goussev:2012,Gorin2006LEreview},  is comprised by the GAAC. This last connection
is perhaps unsurprising as the Loschmidt echo is indeed a measure of auto-correlation of a  dynamicaly evolving state which is precisely
what $1-G_{\cal A}(U)$ does at the more general algebra level. 

The special results {\bf{1}}--{\bf{5}} reported here below can be   obtained by  Eqs. (\ref{eq:Omega-picture}) and (\ref{2-point-GAAC}) by rather straightforward manipulations.

%The fact that average bipartite OTC (i.e., operator entanglement), CGP and Loschmidt echo can be seen as different manifestations
%of the same underlying concept i.e., the GAAC, is an elegant unification and provides one of the main motivations for this work.
%We notice that all these tools have proven valuable in  the study of quantum chaos.

%\subsection{The Bipartite Case}
%\label{sec:bipartite}
%%%%%%%%%%%%%%%%%%%%%%%%%% AVERAGED OTOC

 {\bf{1}}) Now we consider a bipartite quantum system with ${\cal H}= {\cal H}_A\otimes {\cal H}_B$ and ${\cal A}=L({\cal H}_A)\otimes \openone_B$ and, therefore, 
 ${\cal A}^\prime=\openone_A\otimes L({\cal H}_B).$ In this case one finds that 
$\mathbb{P}_{{\cal A}^\prime}(X)=\frac{\openone}{d_A}\otimes\mathrm{Tr}_A(X),\,$
 $\Omega_{\cal A}=\frac{S_{AA^\prime}} {d_A},$ where $S_{AA^\prime}$ is the swap 
 between the $A$ factors in ${\cal H}^{\otimes\,2}$ and $d_X=\mathrm{dim}\,{\cal H}_X\,(X=A,B).$ One gets
 \begin{align}\label{eq:op-ent}
 G_{\cal A}(U)=1-\frac{1}{d^2}\langle  S_{AA^\prime},\,{\cal U}^{\otimes\,2}(S_{AA^\prime})\rangle,
 \end{align}
 where $d=d_B d_A=\mathrm{dim}\,{\cal H}.$
The same relation is true with $S_{AA^\prime}\rightarrow S_{BB^\prime}=S S_{AA^\prime}=d_A\tilde{\Omega}_{\cal A}.$

Eq.~(\ref{eq:op-ent}) \emph{ coincides exactly } with the averaged OTOC discussed in \cite{styliaris_information_2020} i.e.,
$d^{-1} \mathbb{E}_{X\in{\cal A}, Y\in{\cal A}^\prime}\left[\|[X,\,{\cal U}(Y)]\|_2^2\right]$ 
(here $\mathbb{E}$ denotes  the Haar average over the unitary groups of $\cal A$ and ${\cal A}^\prime.$).
Remarkably, this quantity was shown to be  equal to the \emph{operator entanglement} \cite{zanardi2001entanglement, wang2002entanglement} of the unitary $U.$

The latter concept has found important applications to a variety of quantum information-theoretic problems 
\cite{Prosen_operator_2007,chen_operator_2018,alba_operator_2019,bertini_operator_2020-1,bertini_operator_2020-2}.
More recently, it has been shown that operator entanglement 
requires \emph{exponentially scaled} computational resources to simulate \cite{Google-scrambling-2021}. 

{\em{Remark.--}} The  bi-partite OTOC Eq.~(\ref{eq:op-ent}), because of the averages over the two full sub-algebras, does not satisfy Lieb-Robinson type of bounds with associated effective `light-cone" structures. Indeed the regions $A$ and $B$ are complementary and therefore contiguous (zero distance). The same is, in general true, for the GAAC which  does not even require a locality (tensor product) structure to begin with.

%%%%%%%%%%%%%%%%%%%%CGP

{\bf{2)}}  Let ${\cal A}_B$ the algebra of operators which are diagonal with respect to an orthonormal  basis $B:=\{|i\rangle\}_{i=1}^d$ i.e., ${\cal A}_B=\mathbf{C}\,\{ \Pi_i:=|i\rangle\langle i|\}_{i=1}^d.$
This is a $d$-dimensional maximal abelian subalgebra of $L({\cal H})$ such that ${\cal A}={\cal A}^\prime.$ %Notice that this is a special case of the SS algebras discussed in the above.
In this case $\mathbb{P}_{{\cal A}^\prime}(X)=\sum_{i=1}^d \Pi_iX\Pi_i,\,\Omega_{\cal A}=\sum_{i=1}^d\Pi_i^{\otimes\,2},$ and
\begin{align}\label{eq:MASA}
G_{{\cal A}_B}(U)=1-\frac{1}{d}\sum_{i,j=1}^d|\langle i|U|j\rangle|^4,
\end{align}
This expression coincides with the \emph{coherence generating power} (CGP) of $U$ introduced in  \cite{zanardiCoherencegeneratingPowerQuantum2017,styliaris_quantum_2019-1}. 
CGP is there defined as the average coherence (measured
by the  the sum of the square of off-diagonal elements, with respect $B$) generated by $U$ starting from any of the pure incoherent states $\Pi$ i.e., 
$G_{{\cal A}_B}(U)= \frac{1}{d}\sum_{i=1}^d \| \mathbb{Q}_B {\cal U}(\Pi_i)\|_2^2,$  where $\mathbb{Q}=1-\mathbb{P}_{{\cal A}_B}$ projects onto the orthogonal complement of ${\cal A}_B.$ \cite{zanardiCoherencegeneratingPowerQuantum2017, zanardi-CGP-measures-2007}. 
The fact that the CGP is  related to the distance between maximal abelian subalgebras was already established in \cite{zanardiQuantumCoherenceGenerating2018}. 
%$\frac{1}{d}\sum_{i=1}^d\|(1-\mathbb{P}_{{\cal A}_B}) {\cal U}(\Pi_i)\|_2^2.$
CGP  has been applied to the detection of the localization transitions in many-body systems \cite{styliaris_quantum_2019-1}, detection of quantum chaos in closed  and open systems \cite{anand2020quantum}. 
%%%%%%%%%%%%%%%%%%%%%%%%%

%\prlsection{ GAAC and geometry}%%%%%%%%%%%%%%%%%%%%%%% GEOMETRY %%%%%%%%%%%%%%%%%%%%  %(see Remarks above)

{\bf{3)}} ${\cal H}=\mathbf{C}^d\otimes \mathbf{C}^d,$ ${\cal A}=L({\cal H})_{s},$ ${\cal A}^\prime=\mathbf{C}\{\openone,\, S\}\cong\mathbf{C}\mathbf{Z}_2.$
Here, $L({\cal H})_{s}$ denotes the algebra of symmetric operators  i.e., commuting with the swap $S.$ One can readily check that 
$\tilde{\Omega}_{\cal A}= \frac{1}{2}\sum_{J=\pm 1}\left(\frac{\openone+J S}{\sqrt{d(d+J)}}\right)^{\otimes\,2},$ and 
 \begin{align}\label{eq:Sym}
G_{L({\cal H})_{s}}(U)=\frac{1}{2} \left( 1 -\left|\frac{1-\langle S,\,{\cal U}(S)\rangle }{d^2-1}\right|^2 \right)
\end{align}
Here $J=\pm 1$ is labeling the symmetric/anti-symmetric  representation of the permutation group generated by $S.$

{\bf{4)}}
${\cal H}=\mathbf{C}^d\otimes \mathbf{C}^d,$ ${\cal A}=\mathbf{C}\mathbf{Z}_2$  and ${\cal A}^\prime=L({\cal H})_{s}.$ 
Here, $\Omega_{\cal A}=\frac{1}{2}\left( \openone^{\otimes\,2} + S^{\otimes\,2} \right),$ $d({\cal A}^\prime)=\mathrm{Tr}\,\Omega_{\cal A}= d^2(d^2+1)/2,$  whence
%is the algebra of symmetric operators  i.e., commuting with the swap $S.$

\begin{align}\label{eq:Z_2}
G_{\mathbf{C}\mathbf{Z}_2}(U)=\frac{1}{2}\frac{d^4-|\langle S,\,{\cal U}(S)\rangle|^2}{d^2(d^2+1)}
\end{align}
Consistently with Prop.~(\ref{th:Algebra-invariance}) both functions vanish iff  $\langle S,\,{\cal U}(S)\rangle=d^2\Leftrightarrow {\cal U}(S)=S.$
That is to say that non-scrambling $U$ are such that $[U,\,S]=0$ i.e., $U\in L({\cal H})_{s}.$

{\bf{5)}} Let $|\psi\rangle\in{\cal H}$ and $\Pi=|\psi\rangle\langle\psi|.$
%This last example shows in which sense the concept of \emph{Loschmidt echo} ${\cal L}:=|\langle\psi|U|\psi\rangle|$ is encompassed by our formalism. 
We define ${\cal A}_{LE}=\mathbf{C}\{\openone,\,\Pi\}$ i.e., the unital *-closed algebra generated by the projection $\Pi.$
The commutant ${\cal A}_{LE}^\prime$ is the algebra of operators leaving the subspace $\mathbf{C}|\psi\rangle$ and its orthogonal complement invariant. 
One has, $\Omega_{{\cal A}_{LE}}=\Pi^{\otimes\,2}+(\openone-\Pi)^{\otimes\,2},$ $d({\cal A}_{LE}^\prime)=(d-1)^2+1.$		
\begin{align}\label{eq:LE}
G_{{\cal A}_{LE}}(U)=\frac{2(1-{\cal L}^2)[d-2(1-{\cal L}^2)]}{(d-1)^2+1},%=\frac{2}{d}(1-{\cal L}^2)+O(1/d^2),
%G_{\cal A}(U)=\frac{\delta(U)[d-\delta(U)]}{(d-1)^2+1},%\le 2 \delta(U) %=\delta(U)+O(1/d^2),
\end{align}
 where ${\cal L}:=|\langle\psi|U|\psi\rangle|$  is the Loschmidt echo.
%\begin{align}\label{eq:LE}
%$\delta(U):=\|\Pi-{\cal U}(\Pi)\|_2^2=2\,(1-{\cal L}^2).$ %For large $d$ $G_{\cal A}(U)=\delta(U)/d +O(1/d^2).$
%\end{align}
Notice, $G_{{\cal A}_{LE}}(U)=\frac{2}{d}(1-{\cal L}^2)+O(1/d^2)$ and that $2(1-{\cal L}^2)=\|\Pi-{\cal U}(\Pi)\|_2^2,$
i.e., the distance between the algebras ${\cal A}_{LE}^\prime$ and its image ${\cal U}({\cal A}_{LE}^\prime),$ as captured by the GAAC [see Eq.~(\ref{eq:GAAC-is-distance})], in high dimension is 
directly related to the Hilbert-Schmidt distance between the states $\Pi$ and ${\cal U}(\Pi).$
From Eq.~(\ref{eq:LE}) one can see that the GAAC is a monotonic decreasing function of $\cal L$ for $d>4$ and that 
${\cal L}=1\Rightarrow G_{{\cal A}_{LE}}(U)=0.$ For $d=2$ one is back to {\bf{2)}}.
The case ${\cal L}=0$ corresponds to $U\Pi U^\dagger= \openone-\Pi.$ % which leaves $\cal A$ invariant.}.
 
%Moreover, $G_{\cal A}(U)=0$ iff ${\cal L}=1$ and that $G_{\cal A}(U)$ is maximum for ${\cal L}=0.$

\prlsection{Upper bounds and Expectations}% and Hyper-scrambling}%%%%%%%%%%%%%%%%%%%
What are the bounds to algebra scrambling as measured  by the GAAC? Now we would like to answer this question and to see  whether and how those bounds might  be saturated.
\begin{restatable}{prop}{Upper-Bounds}
\label{th:Upper-bounds} i)
\begin{align}\label{eq:Upper-bounds}
G_{\cal A}(U)\le\mathrm{min}\{1-\frac{1}{d({\cal A})},\,1-\frac{1}{d({\cal A}^\prime)}\}=:G_{UB}({\cal A})
\end{align}
ii) if $d({\cal A}^\prime)\le d({\cal A})$ then the bound above is achieved iff $\mathbb{P}_{{\cal A}^\prime} {\cal U} \mathbb{P}_{{\cal A}^\prime}={\cal T}$
where ${\cal T}\colon X\mapsto \mathrm{Tr}(X)\frac{\openone}{d}.$ %is the depolarizing channel. 
iii)  If ${\cal A}^\prime$ is Abelian the bound $1-\frac{1}{d({\cal A}^\prime)}$ is always achieved.
iv) In the collinear case ii) and iii) above hold true with ${\cal A}\leftrightarrow {\cal A}^\prime.$

%In the collinear case if $\cal A$ is abelian then  the bound $1-\frac{1}{d({\cal A})}$ is always achieved.
\end{restatable}
The  saturation condition $\mathbb{P}_{{\cal A}^\prime} {\cal U} \mathbb{P}_{{\cal A}^\prime}={\cal T}$ is quite transparent and intuitive: 
maximal scrambling  is achieved when, from the point of view of the commutant, the dynamics generated by $\cal U$ is just full depolarization.
Physical degrees of freedom supported in ${\cal A}^\prime$ are, quite properly, fully \emph{scrambled}.

Let us now briefly discuss Prop.~(\ref{th:Upper-bounds}) for the physical cases {\bf{1}}--{\bf{5)}}.
In the bipartite example {\bf{1)}}, if $d_A=d_B,$ then (\ref{eq:Upper-bounds}) is achieved for $U=S$ (swap) \cite{styliaris_information_2020}. In the maximal abelian case {\bf{2)}} the bound $1-d^{-1}$ is  saturated by those $U$'s such that
$|\langle i|U|j\rangle|=d^{-1/2},\,(\forall i,j)$ \cite{zanardiCoherencegeneratingPowerQuantum2017}.
In case {\bf{3)}} the bound $\frac{1}{2}$ is achieved for $\langle S,\,{\cal U}(S)\rangle=1,$
which  amounts to the condition ii). On the other hand, in case {\bf{4)}} from Eq.~(\ref{eq:Z_2}) we see that $\langle S,\,{\cal U}(S)\rangle=0\Rightarrow \mathrm{max}_U\, G_{\mathbf{C}\mathbf{Z}_2}(U)=\frac{1}{2}(1+1/d^2)^{-1}<\frac{1}{2}$ i.e.,
bound (\ref{eq:Upper-bounds}) is \emph{not} always achieved.  {\bf{5)}}  
The bound $1/2$ is achieved for $d=2$ only (${\cal A}^\prime$ is abelian). For $d>4$ the maximun is for ${\cal L}=0$ and it is $O(1/d).$

The next general question that we would like to address is: what is the typical value of the GAAC for generic unitaries?
To answer this question we perform an average of (\ref{eq:Omega-picture}) over random, Haar distributed, unitaries.
\begin{restatable}{prop}{Haar-average}\label{th:Haar-average}
\begin{align}\label{eq:Haar-average}
 i)\qquad \overline{G_{\cal A}(U)}^U=\frac{(d^2-d({\cal A}^\prime))(d({\cal A}^\prime)-1)}{d({\cal A}^\prime )(d^2-1)}
 \end{align}
ii) 
%\begin{align}\label{eq:Haar-Levy}
$\mathbf{Prob}_U\left[| G_{\cal A}(U) -  \overline{G_{\cal A}(U)}^U |\ge \epsilon\right]\le\exp[-\frac{d\epsilon^2}{4K^2}].$

iii) In the collinear case $G_{UB}({\cal A})-\overline{G_{\cal A}(U)}^U=O(1/d)$ and
$\mathbf{Prob}_U\left[ G_{UB}({\cal A}) -  G_{\cal A}(U) \ge d^{-1/3}\right]\le\exp[-\frac{d^{1/3}}{16K^2}].$

In ii) and iii) one can choose   $K\ge 40.$
\end{restatable}
As a sanity check, note that Eq.~(\ref{eq:Haar-average}) implies  that the GAAC of any $U$ vanishes when $d({\cal A}^\prime)=d^2$ i.e., ${\cal A}=\mathbf{C}\openone,$ or  $d({\cal A}^\prime)=1$  i.e., ${\cal A}=L({\cal H}).$ 
In fact in these cases $\cal A$ is obviously invariant under the action of \emph{any} $U.$
Point ii) is a direct application of the Levy Lemma on measure concentration: in high dimension (\ref{eq:Haar-average}) is the \emph{typical} value of the GAAC.
Finally, point iii) shows that, in the collinear case, the average value of the GAAC converges to the bound (\ref{eq:Upper-bounds}) when $d$ grows
and that the GAAC is, with overwhelming probability, close to $G_{UB}({\cal A}).$
%%%%%%%%%%%%%%%%%%%%%%%%%%%%%%%%%%%%%%%%%%%%%%%%%%%%%%%%

\prlsection{Time dynamics: infinite averages and fluctuations}
In this final section we will consider a one-parameter group of unitary channels $\{U_t:=e^{-iHt}\}_t$ generated by an Hamiltonian $H$. The idea is that the behavior of the infinite-time average
$\overline{G_{\cal A}(U_t)}^t:=\lim_{T\to\infty} T^{-1}\int_0^T dt\,G_{\cal A}(U_t)$ contains information about the ``chaoticity" of the dynamics as seen from the physical degrees of freedom in the algebra.
These calculations greatly extends the corresponding results, for the bipartite averaged OTOC, reported in \cite{styliaris_information_2020}.
%
%The first observation  in that the, in the light of Eq.~(\ref{eq:Omega-picture}), one can write $\overline{G_{\cal A}(U_t)}^t=1-\|{\cal P}(\Omega_{\cal A})\|_2^2\|\Omega_{\cal A}\|_2^{-2},$
%where ${\cal P}:=\overline{{\cal U}_t^{\otimes\,2}}^t$ is the projector over the commutant of the doubled Hamiltonian $H^{(2)}=H\otimes\openone+\openone\otimes H,$ (which generates ${\cal U}_t^{\otimes\,2}$).
%This cummutant is generated by the spectral projections of $H^{(2)}$ and necessarily always contains $\openone^{\otimes\,2}$ and the swap operator $S.$ Moreover, if $\{|\Psi_k\rangle\}_{k=1}^d$ is \emph{any} basis of eigenvectors
%of $H$ then all projections $\Pi_k\otimes\Pi_l:=|\Psi_k\rangle\langle\Psi_k|\otimes |\Psi_l\rangle\langle\Psi_l|$ are in the commutant as well.
%Using these nested algebras one can prove the following:
%
\begin{restatable}{prop}{Haar-average}\label{th:Time-average}
$\overline{G_{\cal A}(U_t)}^t\le \overline{G_{\cal A}(U_t)}^{NRC}\le \overline{G_{\cal A}(U)}^U$ where
\begin{align}\label{eq:Time-average}
\overline{G_{\cal A}(U_t)}^{NRC}=1-\frac{1}{d({\cal A}^\prime)} \sum_{\alpha=0,1}\left[  \|R^{(\alpha)}\|_2^2  -\frac{1}{2}   \|R_D^{(\alpha)}\|_2^2 \right],
\end{align}
 $R^{(0)}_{lk}:=\| \mathbb{P}_{{\cal A}^\prime}(|\Psi_l\rangle\langle\Psi_k|)\|_2^2,$ $R^{(1)}_{lk}:=\langle \mathbb{P}_{{\cal A}^\prime}(\Pi_l),\, \mathbb{P}_{{\cal A}^\prime}(\Pi_k)\rangle,$
and $(R_D^{(\alpha)})_{lk}:= \delta_{lk} R^{(\alpha)}_{lk},$ $(l,k=1,\ldots, d).$
Moreover, the first inequality above becomes an equality if $H$ fulfills the so-called Non Resonance Condition (NRC).
\end{restatable}
\emph{Remark.--}
The NRC condition amounts to  to say $E_l+E_k=E_n+E_m$ iff $l=n,\,k=m$ or $l=m,\,k=n$
In words: the Hamiltonian spectrum and its gaps are non-degenerate. This fact holds  true for \emph{generic} (non-interacting) Hamiltonians.

The result above, which holds for any observable algebra $\cal A,$ has the very same structure of the corresponding one proved for the averaged bipartite OTOC (see Prop.~4 in \cite{styliaris_information_2020}).
The matrices $R^{(\alpha)},\,(\alpha=0,1)$ encode the connection between the algebra and the \emph{full} system of eigenstates of $H.$

A further simplification occurs, as usual, for the collinear situation $\mathbf{d}=\lambda \mathbf{n}$: $\lambda R^{(0)}_{lk}= \langle \mathbb{P}_{{\cal A}}(\Pi_l),\, \mathbb{P}_{{\cal A}}(\Pi_k)\rangle=:
R^{(1)}_{lk}({\cal A}^\prime).$
%In this case both matrices $R^{(0)})$ and $R^{(1)}$ are bistochastic and, therefore, their $2$-norms are lower bounded by $1.$  It follows that in the collinear
In this  case Eq.~(\ref{eq:Time-average}) can be written in way in which $\cal A$ and ${\cal A}^\prime$ appear symmetrically and 
the following upper bound holds:
\begin{align}\label{eq:UB-NRC}
\overline{G_{\cal A}(U_t)}^{NRC}\le 1-\frac{1}{d({\cal A}^\prime)}-\frac{1}{d({\cal A})} +\frac{1}{d\,d({\cal A}^\prime)}.
\end{align}
This bound is saturated iff $\mathbb{P}_{{\cal A}^\prime}(\Pi_l)=\mathbb{P}_{{\cal A}}(\Pi_l)=\frac{\openone}{d},\,(\forall l).$
Namely,  Hamiltonians whose eigenstates are fully scrambled by the two algebra projections   correspond to maximal infinite-time averaged GAAC.  For these Hamiltonians
infinite-time averages of arbitrary observables are, from  the point of view of  $\cal A$ and ${\cal A}^\prime,$ \emph{completely} randomized \footnote{Indeed, for any 
observable $\mathbb{P}_{\cal A}(\overline{A(t)}^t)= \mathbb{P}_{\cal A}(\sum_l A_l \Pi_l)=
\sum_l A_l  \mathbb{P}_{\cal A}(\Pi_l)=\frac{\openone}{d}\sum_l A_l= \frac{\mathrm{Tr}(A)\openone}{d}={\cal T}(A)$ where $A_l:= \mathrm{Tr}(A\Pi_l)$. Same holds for ${\cal A}^\prime.$}.
Conceptually, this seems a natural way of characterizing  chaoticity relative to the distinguished algebra   of observables.

For example:
In the bipartite case {\bf{1)}} with $d_A=d_B$ the bound (\ref{eq:UB-NRC}) is achieved if the (non-degenerate) Hamiltonian has a fully-entangled eigenstates \cite{styliaris_information_2020}.
In the maximal abelian algebra case {\bf{2)}}  the bound saturation corresponds to Hamiltonians with eigenstates that have maximum coherence with respect to the basis associated with ${\cal A}$
\cite{zanardiCoherencegeneratingPowerQuantum2017}. 
In both these two important physical situations, the RHS of Eq.~(\ref{eq:UB-NRC})  is equal to $(1-\frac{1}{d})^2$ and 
$  \overline{G_{\cal A}(U)}^U - \overline{G_{\cal A}(U_t)}^{NRC}=O(\frac{1}{d^2});$ 
whereby, assuming that NRC holds, using  iii) in Prop.~(\ref{th:Haar-average}) and the  Markov inequality, one can bound temporal fluctuations:
\begin{align}\label{eq:temp-fluct}
\mathbf{Prob}_t\left[ G_{UB}({\cal A})- G_{\cal A}(U_t)\ge \epsilon \right]\le O(\frac{1}{d\,\epsilon}),
\end{align}
one sees e.g., by choosing  $\epsilon=d^{-1/3},$ that Hamiltonians achieving bound (\ref{eq:UB-NRC}) have, in high dimension, highly suppressed temporal fluctuations below the value (\ref{eq:Upper-bounds}).

In \cite{styliaris_information_2020} this concentration phenomenon has been numerically observed for the bi-partite case in chaotic many-body systems and \emph{not} in integrable systems.
For the same type of physical systems, suppression of temporal variance of CGP has been noticed in \cite{anand2020quantum}. These findings were used to suggest that both the bi-partite averaged OTOC and CGP
can be used as diagnostic tools to detect some aspects of quantum chaotic behavior. 
The results above show how this picture may extend to the general algebraic setting developed in this paper.

In fact,we would like to  {\emph{define}}  ${\cal A}$-chaotic the dynamics generated by $U_t$'s  such that
the (relative) difference between its infinite-time average and the Haar-average  of the GAAC is approaching zero sufficiently fast as the system dimension grows.
More formally,   
\begin{align}\label{eq:A-chaos}
 1-\overline{G_{\cal A}(U_t)}^t/ \overline{G_{\cal A}(U)}^{U}=:\epsilon=O(d^{-\gamma})\quad (\gamma\ge 1) .
\end{align}
%Numerical results for many-body integrable chaotic systems in \cite{styliaris_information_2020} provides some support to this characterization for the bi-partite case {\bf{1)}}.
In particular,  in the collinear case, this condition would allow one to prove the ``equilibration" result for the GAAC (\ref{eq:temp-fluct}). 
The intuition behind this definition is quite simple: if Eq.~(\ref{eq:A-chaos}) holds the long time behavior of the GAAC gets, as the system dimension grows, quickly indistinguishable
from the one of a typical Haar random unitary i.e.,  a ``fully chaotic" one.

Before concluding, we would like to  illustrate $\cal A$-chaos with the simple Loschmidt case {\bf{5)}}. Here one has  $\epsilon =\overline{{\cal L}^2_t}^t+O(1/d)$ where
${\cal L}_t=|\langle\psi|U_t|\psi\rangle|.$
The infinite-time average is given by  the purity of the Hamiltonian dephased state $\overline{{\cal L}^2_t}^t=\|\overline{{\cal U}_t(|\psi\rangle\langle\psi|)}^t\|_2^2$ \cite{venuti2010unitary}
%$\overline{{\cal U}_t(|\psi\rangle\langle\psi|)}^t=\sum_l\Pi
%where $D_H$ is the dephasing superoperator associated by the Hamiltonian i.e., $D_H(X):=\lim_{T\to\infty}T^{-1} \int_0^T {\cal U}_t(X),\,(U_t=e^{-i Ht}).$ 
Whence the "chaoticity" condition is achieved if this purity is $O(1/d)$ which in turn implies that the dephased state  is $O(1/d)$ away from  the maximally mixed state. This condition is known to be a sufficient one to bound time-fluctuations of the expectation value of observables with initial state $|\psi\rangle$ \cite{reimann_foundation_2008}. Namely, ${\cal A}_{LE}$-chaos amounts to temporal-equilibration \cite{venuti2010unitary}.

\prlsection{Conclusions}
In this paper we have proposed a novel approach to quantum scrambling based on algebras of observables. A quantitative measure of scrambling 
is introduced in terms of  anti-correlation between the whole commutant algebra and its (unitarily) evolved image. 
This quantity, which we named the Geometric Algebra Anti Correlator (GAAC),
has also a clear geometrical meaning as it describes the distance between the two algebras or, equivalently, the degree of self-orthogonalization induced by the dynamics. 

We explicitly computed the GAAC for several physically motivated cases and characterized its behavior in terms of typical values, upper bounds and temporal fluctuations.
 We have shown that  the GAAC formalism provides an unified mathematical and conceptual framework for  concepts like operator entanglement, averaged bipartite OTOC, coherence generating power and Loschmidt echo.

Finally, we suggested an approach to quantum chaos in terms of the behavior of infinite-time average of the GAAC for large system dimension.
To assess the  effectiveness of such an  approach is one of the challenges of future investigations.

%%%%%%%%%%%%%%%%%%%%%%%%%%%%%%%%%%%%%%%%%%%%%%%%%%%%%%%%
%%%%%%%%%%%%%%%%%%%%%%%%%%%%%%%%%%%%%%%%%%%%%%%%%%%%%%%%%%%%%%%%%%%%%%%%%%%%%%%%%%%%%%%%%%%%%%%%%%%%%%%%%%%%%%%%%%%%%%%%%%%%%%%%%%%%%%%%%%%%%%%%%%%%%%%%%%%%%%%%%%%%%%%%
%%%%%%%%%%%%%%%%%%%%%%%%%%%%%%%%%%%%%%%%%%%%%%%%%%%%%%%%%%%%%%%%%%%%%%%%%%%%%%%%%%%%%%%%%%%%%%%%%%%%%%%%%%%%%%%%
%%%%%%%%%%%%%%%%%%%%%%%%%%%%%%%%%%%%%%%%%%%%%%%%%%%%%%%%%%%%%%%%%%%%%%%%%%%%%%%%%%%%%%%%%%%%%%%%%%%%%%%%%%%%%%%%%%%%%%%%%%%%%%%%%%%%%%%%%%%%%%%%%%%%%%%%%%%%%%%%%%%%%%%%
%%%%%%%%%%%%%%%%%%%%%%%%%%%%%%%%%%%%%%%%%%%%%%%%%%%%%%%%%%%%%%%%%%%%%%%%%%%%%%%%%%%%%%%%%%%%%%%%%%%%%%%%%%%%%%%%
%%%%%%%%%%%%%%%%%%%%%%%%%%%%%%%%%%%%%%%%%%%%%%%%%%%%%%%%%%%%%%%%%%%%%%%%%%%%%%%%%%%%%%%%%%%%%%%%%%%%%%%%%%%%%%%%%%%%%%%%%%%%%%%%%%%%%%%%%%%%%%%%%%%%%%%%%%%%%%%%%%%%%%%%
%%%%%%%%%%%%%%%%%%%%%%%%%%%%%%%%%%%%%%%%%%%%%%%%%%%%%%%%
\section{Acknowledgments}
I acknowledge discussions with Namit Anand and partial support from the NSF award PHY-1819189. This research was (partially) sponsored by the Army Research Office and was accomplished under Grant Number W911NF-20-1-0075. The views and conclusions contained in this document are those of the authors and should not be interpreted as representing the official policies, either expressed or implied, of the Army Research Office or the U.S. Government. The U.S. Government is authorized to reproduce and distribute reprints for Government purposes notwithstanding any copyright notation herein.
\bibliographystyle{apsrev4-1}
\bibliography{refs,my_library}
\appendix
\begin{widetext}
\section{Supplemental Material}
\subsection{Proof of Prop \ref{th:Algebra-invariance}}
\emph{i)} It is a direct computation: $D^2({\cal A}^\prime,\, {\cal U}({\cal A}^\prime))=\|\mathbb{P}_{{\cal A}^\prime}-\mathbb{P}_{{\cal U}({\cal A}^\prime)}\|_{HS}^2= \|\mathbb{P}_{{\cal A}^\prime}\|_{HS}^2+\|\mathbb{P}_{{\cal U}({\cal A}^\prime)}\|_{HS}^2-
2\,\langle \mathbb{P}_{{\cal A}^\prime},\, \mathbb{P}_{{\cal U}({\cal A}^\prime)} \rangle.$ Now, $ \|\mathbb{P}_{{\cal A}^\prime}\|_{HS}^2=\|{\cal U} \mathbb{P}_{{\cal A}^\prime}{\cal U}^\dagger\|_{HS}^2= \|\mathbb{P}_{{\cal U}({\cal A}^\prime)}\|_{HS}^2=d({\cal A}^\prime).$ Whence, by dividing $D^2$ by $2\,d({\cal A}^\prime)=2\,  \|\mathbb{P}_{{\cal A}^\prime}\|_{HS}^2,$ Eq.~(\ref{eq:GAAC-is-distance}) follows.

\emph{ii)} Since $D$ is a metric from Eq.~(\ref{eq:GAAC-is-distance}) one has $G_{\cal A}(U)=0\Leftrightarrow {\cal U}({\cal A}^\prime)={\cal A}^\prime\Leftrightarrow 
{\cal U}({\cal A})={\cal A}.$ Last equivalence is obtained by taking the commutant of both sides, using ${\cal A}^{\prime\prime}={\cal A}$ (double  commutant theorem) and that ${\cal U}({\cal A})^\prime= {\cal U}({\cal A}^\prime)$ (true for unitary auto-morphisms).
%%%%%%%%%%%%%%%%%%%%%%%%%%%%%%%%%%%%%%%%%%%%%%%%%%%
\subsection{Proof of Prop \ref{th:Omega-picture}}
\emph{i)} Let us write the Algebra projections in the Kraus form $\mathbb{P}_{{\cal A}^\prime}(X)=\sum_\alpha e_\alpha  X e^\dagger_\alpha $ and $\mathbb{P}_{{\cal U}({\cal A}^\prime)}(X)=
\sum_\alpha (U e_\alpha U^\dagger)   X  (U e_\alpha U^\dagger)^\dagger. $ 
Here, because of the structure theorem (\ref{eq:Alg-central-decomposition}), one can choose the \emph{orthogonal}  basis of ${\cal A}$ given by
\begin{align}\label{eq:e_a}
e_\alpha=\frac{1}{\sqrt{d_J}} \openone_{n_J}\otimes |l\rangle\langle m|,\in{\cal A}\qquad \alpha:=(J, l, m)\, (l,m=1,\ldots, d_J).
\end{align}
(note that $|\{\alpha\}|=\sum_Jd_J^2=d({\cal A})$ and that the set $\{e_\alpha\}_\alpha$ is closed under hermitian conjugation).
Hence $\langle \mathbb{P}_{{\cal A}^\prime},\, \mathbb{P}_{{\cal U}({\cal A}^\prime)} \rangle=
\mathrm{Tr}_{HS}\left( \mathbb{P}_{{\cal A}^\prime} \mathbb{P}_{{\cal U}({\cal A}^\prime)}\right)=\sum_{\alpha,\beta} |\mathrm{Tr}\left(e_\alpha U  e_\beta U^\dagger     \right)   |^2=
\sum_{\alpha,\beta} |\langle e_\alpha,\,{\cal U}(e_\beta)\rangle|^2.$ Here we've used that  ${\cal T}=\sum_i T_i XT_i^\dagger\Rightarrow\mathrm{Tr}_{HS}\,{\cal T}=\sum_i |\mathrm{tr}\, T_i|^2.$
Because of the definition (\ref{eq:GAAC}) this proves Eq.~(\ref{2-point-GAAC}).

On the other hand, if $\Omega_{\cal A}=\sum_\alpha e_\alpha\otimes e_\alpha^\dagger$ one has $\langle \Omega_{\cal A},\,{\cal U}^{\otimes\,2}(\Omega_{\cal A})\rangle=\sum_{\alpha,\beta}
\langle e_\alpha\otimes e_\alpha^\dagger,\,  U  e_\beta U^\dagger \otimes U  e^\dagger_\beta U^\dagger \rangle=\sum_{\alpha,\beta} \langle e_\alpha,\, U  e_\beta U^\dagger\rangle
 \langle e^\dagger_\alpha,\, U  e^\dagger_\beta U^\dagger\rangle= \sum_{\alpha,\beta} |\langle e_\alpha,\,{\cal U}(e_\beta)\rangle|^2.$ Moreover, from (\ref{eq:e_a}) one has
 \begin{align}\label{eq;Omega_A}
 \Omega_{\cal A} =\sum_J \frac{\openone_{n_J}^{\otimes\,2}}{d_J}\otimes \sum_{l,m=1}^{d_J} |lm\rangle\langle ml|=:\sum_J \frac{\openone_{n_J}^{\otimes\,2}\otimes S_{d_J}}{d_J}.
 \end{align}
 Therefore, $\|\Omega_{\cal A} \|_2^2=\mathrm{Tr}\left(  \Omega^2_{\cal A} \right)= \mathrm{Tr}\left(  \sum_J \frac{(\openone_{n_J}\otimes \openone_{d_J})^{\otimes\,2}}{d_J^2} \right)=\sum_J (n_J d_J)^2/d_J^2=\sum_J n_J^2=d({\cal A}^\prime).$ This completes the proof of the first equality in Eq.~(\ref{eq:Omega-picture}). Now, if $S$ is the swap operator
 $S \Omega_{\cal A} S=\sum_\alpha e_\alpha^\dagger \otimes e_\alpha=\sum_\alpha  e_\alpha\otimes e_\alpha^\dagger=\Omega_{\cal A},$ i.e.,$[S,\,\Omega_{\cal A}]=0.$
 Since $[U^{\otimes\,2},\,S]=0$ and $\|\Omega_{\cal A}\|_2^2=\|S\Omega_{\cal A}\|_2^2=\|{\tilde{\Omega}_{\cal A}}\|_2^2$ the second equality in Eq.~(\ref{eq:Omega-picture}) follows.
 Also, since $S(\openone_{n_J}^{\otimes\,2}\otimes S_{d_J})=S_{n_J}\otimes  \openone_{d_J}^{\otimes\,2},$ ($S_{n_J}$ is a swap operator defined over the $\mathbf{C}^{n_J}$ factors) one finds
 $\tilde{\Omega}_{\cal A}=:\sum_J \frac {S_{n_J}\otimes \openone_{d_J}^{\otimes\,2}}{d_J}=\sum_\gamma f_\gamma\otimes f^\dagger_\gamma,$
 where $\gamma:=(J, p,q),\,p,q=1,\ldots,n_J$ and $f_\gamma= \frac{1}{\sqrt{d_J}}|p \rangle\langle q|\otimes \openone_{d_J}\in{\cal A}^\prime.$
 This is an \emph{orthonormal} basis of ${\cal A}^\prime.$

By direct computation $\mathrm{Tr}_1\left( S \Omega_{\cal A} (X\otimes\openone\right)=\sum_J \mathrm{tr}_{d_J}(X)\otimes \frac{\openone_{d_J}}{d_J}=\mathbb{P}_{{\cal A}^\prime}(X)$
which proves Eq.~(\ref{eq:Omega-projections}).

\emph{ii)}
In the collinear case $\tilde{\Omega}_{\cal A}=\frac{1}{\lambda} {\Omega}_{{\cal A}^\prime}= \frac{1}{\lambda}  \sum_J \frac {S_{n_J}\otimes \openone_{d_J}^{\otimes\,2}}{n_J}$
where $d_J=\lambda n_J,\,(\forall J).$  Inserting this in Eq.~(\ref{eq:Omega-picture}) and using $ \lambda^2 \|{\Omega}_{\cal A}\|_2^2=  \lambda^2 \sum_J n_J^2= \sum_Jd_J^2=d({\cal A})=
\|  {\Omega}_{{\cal A}^\prime}\|_2^2$ one sees that  in this collinear case $G_{\cal A}(U)=G_{{\cal A}^\prime}(U).$

%%%%%%%%%%%%%%%%%%%%%%%%%%%%%%%%%%%%%%%%%%%%%%%%%%%%%%%%%%%%%%%%%%
\subsection{Proof of Prop \ref{th:Upper-bounds}}
\emph{i)} First, notice that for any two orthogonal projections $P$ and $Q$ one has that $\mathrm{Tr}\left(PQ\right)\ge \mathrm{dim}\left(   \mathrm{Im} P\cap   \mathrm{Im} Q    \right).$
Since both ${\cal A}^\prime$ and ${\cal U}({\cal A}^\prime)$ contain the identity $\openone$ one has $\langle \mathbb{P}_{{\cal A}^\prime},\, \mathbb{P}_{{\cal U}({\cal A}^\prime)}\rangle\ge 1,$
from which the bound $G_{\cal A}(U)\le 1-1/d({\cal A}^\prime),$ immediately follows.

To prove the bound $G_{\cal A}(U)\le 1-1/d({\cal A}),$ we begin by observing that ${\Omega}_{\cal A}=\sum_\alpha \tilde{e}_\alpha\otimes  \tilde{e}^\dagger_\alpha$
for any basis  $\tilde{e}_\alpha=\sum_\beta U_{\beta,\alpha} e_\beta,$ where the $e_\beta$'s are given by (\ref{eq:e_a}) and the matrix $U_{\beta,\alpha}$ is unitary. 
Now $\frac{\openone}{\sqrt{d({\cal A})}}=\sum_J \sqrt{\frac{d_J}{d({\cal A})}}\sum_{l=1}^{d_J} e_{(J,l,l)}=:\tilde{e}_1.$
Since $\sum_{J,l} |\sqrt{\frac{d_J}{d({\cal A})}}|^2= \frac{1}{d({\cal A})}\sum_J d_J^2=1,$ we see that on can always unitarily move to a new basis such that
$\tilde{e}_1=\frac{\openone}{\sqrt{d({\cal A})}}.$ Whence ${\Omega}_{\cal A}= \frac{\openone}{{d({\cal A})}}+\sum_{\alpha>1} \tilde{e}_\alpha\otimes  \tilde{e}^\dagger_\alpha=:
\frac{\openone}{{d({\cal A})}}+\Omega_{\cal A}^\prime$ and  $\langle \Omega_{\cal A},\,{\cal U}^{\otimes\,2}(\Omega_{\cal A})\rangle=\frac{\mathrm{Tr}\, \Omega_{\cal A} }{d({\cal A})} +
\langle \Omega_{\cal A},\,{\cal U}^{\otimes\,2}(\Omega^\prime_{\cal A})\rangle\ge\frac{d({\cal A}^\prime)}{d({\cal A})}$ ( note $\langle \Omega_{\cal A},\,{\cal U}^{\otimes\,2}(\Omega^\prime_{\cal A})\rangle
=\sum_{\alpha}\sum_{\beta>1} |\langle e_\alpha,\,{\cal U}(\tilde{e}_\beta)\rangle|^2\ge 0.$). Plugging this inequality in Eq.~(\ref{eq:Omega-picture}) one finds $G_{\cal A}(U)\le 1-1/d({\cal A}).$

In summary, $G_{\cal A}(U)\le \mathrm{min}\{ 1-1/d({\cal A}),\,1-1/d({\cal A}^\prime)\}.$

\emph{ii)} One has that $\langle \mathbb{P}_{{\cal A}^\prime},\, \mathbb{P}_{{\cal U}({\cal A}^\prime)}\rangle=\| \mathbb{P}_{{\cal A}^\prime} {\cal U} \mathbb{P}_{{\cal A}^\prime}  \|_{HS}^2.$
This last norm is always larger than the (square of the) operator norm of the CP-map ${\cal F}:= \mathbb{P}_{{\cal A}^\prime} {\cal U} \mathbb{P}_{{\cal A}^\prime}$ which is one.
The lower bound is achieved \emph{iff} ${\cal F}$ has rank one, but the only rank one unital trace-preserving CP-map is the depolarizing channel ${\cal T}.$
 
 \emph{iii)} If ${\cal A}^\prime$ is Abelian one  take the orhonormal basis $f_J =\frac{\openone_{1}\otimes \openone_{d_J}}{\sqrt{d_J}}= \frac{\Pi_J}{\sqrt{d_J}}.$%{\sqrt{d_J}},\,J=1,\ldots,d__Z=d({\cal A}^\prime).$
 It follows, $\langle \tilde{\Omega}_{\cal A},\,{\cal U}^{\otimes\,2}(\tilde{\Omega}_{\cal A})\rangle=\sum_{J,K} \frac{1}{d_J d_K}|\langle \Pi_J,\,{\cal U}(\Pi_K)\rangle|^2.$
 Now if  $\Pi_J=\sum_{l=1}^{d_J} |J l\rangle\langle Jl|$ and $\Pi_K=\sum_{m=1}^{d_K} |Km\rangle\langle K m|,$   
 one has
 $\langle \Pi_J,\,{\cal U}(\Pi_K)\rangle=\sum_{l=1}^{d_J}\sum_{m=1}^{d_K} |\langle Km|U|Jl\rangle|^2.$
 Therefore, if $U$ maps the basis $|Jl\rangle$ into mutually unbiased one i.e., $|\langle Km|U|Jl\rangle|=1/\sqrt{d},\,(\forall J,K,l,m)$
 one finds $\langle \Pi_J,\,{\cal U}(\Pi_K)\rangle=\frac{d_J d_K}{d}$ whence $\sum_{J,K} \frac{1}{d_J d_K}|\langle \Pi_J,\,{\cal U}(\Pi_K)\rangle|^2=\sum_{JK} \frac{d^2_J d^2_K}{d_J d_K d^2}=1$
 (here we used that, in the Abelian case, $\sum_Jd_J=d$.) By Eq.~(\ref{eq:Omega-picture}) this last relation implies that, for these $U$'s, the upper bound $1-1/d({\cal A}^\prime)$ is saturated.
 Notice that in this case, $\mathbb{P}_{{\cal A}^\prime}{\cal U}( \frac{\Pi_J}{\sqrt{d_J}})=\sum_K  \frac{\Pi_K}{\sqrt{d_K}}\langle  \frac{\Pi_K}{\sqrt{d_K}},\,{\cal U}( \frac{\Pi_J}{\sqrt{d_J}})\rangle=
 \frac{1}{d}\sum_K   \frac{\Pi_K}{\sqrt{d_K}} \frac{d_J d_K}{\sqrt{d_J} \sqrt{d_K}}=\sqrt{d_J} \frac{1}{d}\sum_K \Pi_K=\sqrt{d_J}  \frac{\openone}{d}={\cal T}(\frac{\Pi_J}{\sqrt{d_J}}),\,\forall J$
 which implies $\mathbb{P}_{{\cal A}^\prime}{\cal U}\mathbb{P}_{{\cal A}^\prime}={\cal T}.$ 
 
\emph{iv)} Since in the collinear case $G_{\cal A}(U)=G_{{\cal A}^\prime}(U)$ ii) and iii) above holds with $\cal A$ replacing ${\cal A}^\prime.$
%%%%%%%%%%%%%%%%%%%%%%%%%%%%%%%%%%%%%%%%%%%%%%%%%%%%%%%%%%%%%%%%%%
\subsection{Proof of Prop \ref{th:Haar-average}}
\emph{i)} Averaging over the Haar measure gives you a projector: $ \overline{{\cal U}^{\otimes\,2}(X)}^U= \overline{  U^{\otimes\,2} X  U^{\dagger\otimes\,2}}^U=:\mathbb{P}_{Haar}(X)$ over the commutant of the
algebra generated by $\{U^{\otimes\,2}\,/\, U\in U({\cal H})\}.$ By Schur-Weyl duality this commutant is generated by $\openone$ and the swap $S$:
\begin{align}\label{eq:P_haar}
\mathbb{P}_{Haar}(X)=\frac{1}{2}\sum_{\alpha=\pm 1} \frac{\openone+\alpha S}{d(d+\alpha)}\langle \openone+\alpha S ,\,X\rangle.
\end{align} 
Therefore, $\overline{\langle \Omega_{\cal A},\,{\cal U}^{\otimes\,2}( \Omega_{\cal A})\rangle}^U=\langle \Omega_{\cal A},\,\mathbb{P}_{Haar}( \Omega_{\cal A})\rangle=
\|\mathbb{P}_{Haar}( \Omega_{\cal A})\|_2^2=
\frac{1}{2}\sum_{\alpha=\pm 1} \frac{|\langle \openone+\alpha S ,\,\Omega_{\cal A}\rangle |^2}{d(d+\alpha)}.$
Now $\langle \openone,\,\Omega_{\cal A}\rangle=\mathrm{Tr} \,\Omega_{\cal A}=d({\cal A}^\prime),$ and
$\langle S,\,\Omega_{\cal A}\rangle=\mathrm{Tr}\, \tilde{\Omega}_{\cal A}=\sum_J n_J d_J^2/d_J=\sum_Jn_J d_J=d.$
Proving Eq.~(\ref{eq:Haar-average}) is now straightforward algebra from these equations and (\ref{eq:Omega-picture}).

\emph{ii)} This is an application of the Levy's Lemma for the GAAC: $U\in U({\cal H})\mapsto G_{\cal A}(U):=  \frac{\langle \Omega_{\cal A},\,{\cal U}^{\otimes\,2}(\Omega_{\cal A})\rangle}{\|\Omega_{\cal A}\|_2^2}.$

{\bf{Levy's Lemma}}: 
 \begin{align}|  G_{\cal A}(U) -G_{\cal A}(V)|\le K\|U-V\|_2 \Rightarrow 
\mathbf{Prob}_U\left[ | G_{\cal A}(U)-\overline{G_{\cal A}(U)}^U|\ge\epsilon  \right]\le \exp\left(  -\frac{d\epsilon^2}{4 K^2}  \right).
\end{align}
Let us show that this is Lipschitz function. $| G_{\cal A}(U)- G_{\cal A}(V)|=\|\Omega_{\cal A}\|_2^{-2}| \langle \Omega_{\cal A},\,({\cal U}^{\otimes\,2}- {\cal V}^{\otimes\,2})(\Omega_{\cal A})\rangle|\le 
\|{\cal U}^{\otimes\,2}- {\cal V}^{\otimes\,2}\|_{2,2}$ [here $\|{\cal T}\|_{2,2}:= \sup_{\|X\|_2=1}\|{\cal T}(X)\|_2.$]
If ${\cal U}-{\cal V}=\Delta$ one has $\|{\cal U}^{\otimes\,2}- {\cal V}^{\otimes\,2}\|_{2,2}\le \|\Delta\|_{2,2}( \|\Delta\|_{2,2} +2)$ [where we used $\|X\otimes Y\|_{2,2}=\|X\|_{2,2} \| Y\|_{2,2},$ and $\|{\cal V}\|_{2,2}=1.$]
Moreover, if $U-V=\delta$ then $\Delta(X)=\delta  X \delta^\dagger +\delta X V^\dagger+ V X\delta^\dagger.$ From this one finds
$\|\Delta\|_{2,2}\le  \sup_{\|X\|_2=1} ( \| \delta\|_2 \| X \delta^\dagger\|_2 + \|\delta X\|_2 + \|X\delta^\dagger \|_2)\le 4\,\|\delta\|_2.$
Notice also, $\|X\delta\|_2=\| X (U-V)\|_2\le 2 \|X\|_2,$ whence $\|\Delta\|_{2,2}\le\sup_{\|X\|_2=1} (2\|X\|_2 +2\|X\|_2 +2 \|X\delta^\dagger\|_2)\le 8.$
Bringing everything together: 
\begin{align}
| G_{\cal A}(U)- G_{\cal A}(V)| \le \|\Delta\|_{2,2}( \|\Delta\|_{2,2} +2)\le 10\, \|\Delta\|_{2,2} \le 40 \|\delta\|_2= 40\,\| U-V\|_2.
\end{align}
This shows that one can choose a Lipschitz constant $K\ge 40$ for $f.$ 

\emph{iii)} 
In this collinear  case, since $d^2= d({\cal A})d({\cal A}^\prime),$ the Haar average (\ref{eq:Haar-average}) takes the form $\overline{G_{\cal A}(U)}^U=(1-1/d^2)^{-1} (1-1/d({\cal A}^\prime))(1-1/d({\cal A})).$
Suppose $d({\cal A}^\prime)\le d,$ ($d({\cal A})\ge d$) then $G_{UB}({\cal A})=1-1/d({\cal A}^\prime)$
One has $ G_{UB}({\cal A})- \overline{G_{\cal A}(U)}^U=\frac{1-1/d({\cal A}^\prime)}{1-1/d^2}(1/d({\cal A})-1/d^2)\le 1/d({\cal A})-1/d^2\le 1/d.$
The case $d({\cal A}^\prime)> d,$ works exactly in the same way (with ${\cal A}\leftrightarrow {\cal A}^\prime.$)
This proves that $G_{UB}({\cal A})- \overline{G_{\cal A}(U)}^U=O(1/d).$

Now, $G_{UB}({\cal A}) -G_{\cal A}(U)= (G_{UB}({\cal A}) -\overline{G_{\cal A}(U)}^U) +(\overline{G_{\cal A}(U)}^U-G_{\cal A}(U))\ge \epsilon$
implies, for large $d$, that $\overline{G_{\cal A}(U)}^U -G_{\cal A}(U))\ge \epsilon -1/d\ge \epsilon/2.$
It follows that $\mathbf{Prob}_U\left[   G_{UB}({\cal A}) -G_{\cal A}(U)\ge \epsilon\right] \le \mathbf{Prob}_U\left[ \overline{G_{\cal A}(U)}^U -G_{\cal A}(U))\ge \epsilon/2  \right]\le 
\exp[-\frac{d\epsilon^2}{16 K^2}].$
%%%%%%%%%%%%%%%%%%%%%%%%%%%%%%%%%%%%%%%%%%%%%%%%%%%%%%%%
\subsection{Proof of Prop \ref{th:Time-average}}
Suppose the unitary evolution has the spectral resolution $U_t=\sum_n \Pi_n e^{-iE_n t}$ (here $n$  ranges over the set of distinct eigenvalues) then one has
${\cal U}_t^{\otimes\,2}(X)=\sum_{k,h,p,m} \Pi_k\otimes \Pi_h \,X\, \Pi_p\otimes \Pi_m \,\exp[-it( E_k+E_h-E_p-E_m)].$
Performing the infinite time average  one has  $\overline{\exp[-it( E_k+E_h-E_p-E_m)]}^t=,\delta_{0, E_k+E_h-E_p-E_m}.$ and therefore
\begin{align}\label{eq:P-bar}
\bar{\mathbb{P}}(X):=\overline{{\cal U}_t^{\otimes\,2}(X)}^t= 
 \sum_{k,h,p,m} \Pi_k\otimes \Pi_h \, X\, \Pi_p\otimes \Pi_m \,\delta_{0, E_k+E_h-E_p-E_m}.
 \end{align}
   Clearly, the map $X\mapsto \bar{\mathbb{P}}(X)$ is an orthogonal projection whose range contains $\openone$ and $S$ from which $\bar{\mathbb{P}}\ge {\mathbb{P}}_{Haar}.$
 Limiting the sum to the pairs $k=p,\,h=m$ \emph{or}  $k=m,\,h=p$ (with $k\neq h$) one obtains the sub-projection
 $\mathbb{P}_{NRC^+}\le \bar{\mathbb{P}}.$   By direct inspection $\openone$ and $S$ also belong to the range of $\mathbb{P}_{NRC^+},$ which in turn, implies ${\mathbb{P}}_{NRC^+}\ge {\mathbb{P}}_{Haar}.$
 
A further operator inequality can be obtained considering the (NB non-unique) resolution of the projectors $\Pi_l$'s. In fact, if $\Pi_l=\sum_{j=1}^{d_l} |\psi_{l,j}\rangle\langle \psi_{l,j}|=:\sum _{j=1}^{d_l} \Pi_{l,j}$
one has $\mathbb{P}_{NRC^+}\ge \mathbb{P}_{NRC}$ where
\begin{align}
 \mathbb{P}_{NRC}(X)=\sum_{k,h,i,j}  \Pi_{k,i} \otimes \Pi_{h,j}  \, X\,  \Pi_{k,i} \otimes \Pi_{h,j} +\sum_{k\neq h, i,j}  \Pi_{k,i} \otimes \Pi_{h,j} \,X\,  \Pi_{h,j}\otimes \Pi_{k,i}. 
\end{align}
 This projector equals $\bar{\mathbb{P}}$ if the Non Resonant Condition (NRC) holds i.e., $E_k+E_h=E_p+E_m \Leftrightarrow (k=p) \,\&\, (h=m)$ \emph{or} $(k=m)\, \& \,(h=p)\,\&\,(k\neq h).$ 
  Notice that NRC implies (for $k=h$) that the Hamiltonian is non-degenerate.
  
In summary, so far we have proven that $ \overline{\langle \Omega_{\cal A},\, {\cal U}_t^{\otimes\,2}(\Omega_{\cal A})\rangle}^t= \langle \Omega_{\cal A},\,\bar{\mathbb{P}}(\Omega_{\cal A})\rangle =
\| \bar{\mathbb{P}}(\Omega_{\cal A})\|_2^2 \ge \| {\mathbb{P}}_{NRC^+}(\Omega_{\cal A})\|_2^2 \ge  \| {\mathbb{P}}_{NRC}(\Omega_{\cal A})\|_2^2 \ge  \| {\mathbb{P}}_{Haar}(\Omega_{\cal A})\|_2^2.$ 
These inequalities, by Eq.~(\ref{eq:Omega-picture}), translate into the following ones for the GAAC:
\begin{align} 
\overline{G_{\cal A}(U_t)}^t\le   \overline{G_{\cal A}(U_t)}^{NRC^+}\le   \overline{G_{\cal A}(U_t)}^{NRC}\le  \overline{G_{\cal A}(U)}^U.
\end{align}
 Let us now consider $ \overline{G_{\cal A}(U_t)}^{NRC}$ explicitly. From now on the labels $k,\,h=1,\ldots,d$ correspond to eigenstates (\emph{not} eigenspaces).
 First $  \| {\mathbb{P}}_{NRC}(\Omega_{\cal A})\|^2_2=\sum_{k,h} \|   \Pi_{k} \otimes \Pi_{h}  \, \Omega_{\cal A}\,  \Pi_{k} \otimes \Pi_{h} \|_2^2  +\sum_{k,\neq h} \|  \Pi_{k} \otimes \Pi_{h} \,\Omega_{\cal A}\,  \Pi_{h}\otimes \Pi_{k}  \|_2^2=
 \sum_{k,h} |\langle \psi_k \psi_h|\Omega_{\cal A}| \psi_k \psi_h\rangle|^2 + \sum_{k,\neq h} |\langle \psi_k \psi_h|\Omega_{\cal A}| \psi_h \psi_k\rangle|^2=
\sum_{k,h} |\langle \psi_k \psi_h|\Omega_{\cal A}| \psi_k \psi_h\rangle|^2 + \sum_{k, h} |\langle \psi_k \psi_h|\Omega_{\cal A}| \psi_h \psi_k\rangle|^2  -\sum_{h} |\langle \psi_h \psi_h|\Omega_{\cal A}| \psi_h \psi_h\rangle|^2$
 
 Moreover,
 $
 \langle \psi_k \psi_h|\Omega_{\cal A}| \psi_k \psi_h\rangle=\sum_\alpha \langle \psi_k |e_\alpha|\psi_k\rangle \langle \psi_h |e^\dagger_\alpha|\psi_h\rangle=\sum_\alpha
 \mathrm{Tr}\left(  |\psi_h\rangle\langle\psi_k|e_\alpha|\psi_k\rangle\langle \psi_h |e^\dagger \right)=\langle |\psi_k\rangle  \langle \psi_h |,\,\mathbb{P}_{{\cal A}^\prime}(|\psi_k\rangle  \langle \psi_h |)\rangle=
 \|\mathbb{P}_{{\cal A}^\prime}(|\psi_k\rangle  \langle \psi_h|)\|_2^2=:R^{(0)}_{lk},
 $ and

 $
 \langle \psi_k \psi_h|\Omega_{\cal A}| \psi_h \psi_k\rangle=\sum_\alpha \langle \psi_k |e_\alpha|\psi_h\rangle \langle \psi_h |e^\dagger_\alpha|\psi_k\rangle=\sum_\alpha
 \mathrm{Tr}\left(  |\psi_k\rangle\langle\psi_k|e_\alpha|\psi_h\rangle\langle \psi_h |e^\dagger \right)=\langle |\psi_k\rangle  \langle \psi_k |,\,\mathbb{P}_{{\cal A}^\prime}(|\psi_h\rangle  \langle \psi_h |)\rangle=
 \langle\mathbb{P}_{{\cal A}^\prime}(|\psi_k\rangle  \langle \psi_k |),\, \mathbb{P}_{{\cal A}^\prime}(|\psi_h\rangle  \langle \psi_h |)\rangle=:R^{(1)}_{lk}.
 $
 
 Since, $ \overline{G_{\cal A}(U_t)}^{NRC}=1-d({\cal A}^\prime)^{-1}  \| {\mathbb{P}}_{NRC}(\Omega_{\cal A})\|^2_2,$ the equations above prove Eq.~(\ref{eq:Time-average}).
 
 \prlsection{Collinear case}
 Notice, $\langle \psi_k \psi_h|\Omega_{\cal A}| \psi_k \psi_h\rangle=\langle \psi_k \psi_h|\tilde{\Omega}_{\cal A}| \psi_h \psi_k\rangle$ since in the \emph{collinear} case $\tilde{\Omega}_{\cal A}=\lambda^{-1}\Omega_{{\cal A}^\prime}$
 and $d({\cal A}^\prime)\lambda^2= d({\cal A}),$ from Eq.~(\ref{eq:Time-average}) it follows  that
\begin{align} \label{eq:Time-Average-Collinear}
 \overline{G_{\cal A}(U_t)}^{NRC}=1 -\frac{1}{d({\cal A})}\sum_{l,k}|\langle  \mathbb{P}_{{\cal A}}(\Pi_l),\,  \mathbb{P}_{{\cal A}}(\Pi_k) \rangle|^2-
  \frac{1}{d({\cal A}^\prime)}\sum_{l,k}|\langle  \mathbb{P}_{{\cal A}^\prime}(\Pi_l),\,  \mathbb{P}_{{\cal A}^\prime}(\Pi_k) \rangle|^2+\frac{1}{d({\cal A}^\prime)}\sum_{l}\|\mathbb{P}_{{\cal A}^\prime}(\Pi_l)\|_2^4.
  \end{align}
  The first two terms are basically H-S norms of bistochastic matrices, the third in a diagonal contribution that could also be written as $\frac{1}{d({\cal A})}\sum_{l}\|\mathbb{P}_{{\cal A}}(\Pi_l)\|_2^4.$
  The quantity (\ref{eq:Time-Average-Collinear}) is maximized , providing the upper-bound (\ref{eq:UB-NRC}), when all the entries of both matrices are equal to $1/d$ (maximally mixed bistochastic matrices  with constant entries, a 1D projector). This is true iff
  $\mathbb{P}_{{\cal A}^\prime}(\Pi_l)=\mathbb{P}_{{\cal A}}(\Pi_l)=\frac{\openone}{d},\,(\forall l).$ In terms of the Hamiltonian
  dephasing map  ${\cal D}_H(X)=\sum_l\Pi_l X \Pi_l,$ this maximal scrambling condition reads $\mathbb{P}_{{\cal A}^\prime} {\cal D}_H=
  \mathbb{P}_{{\cal A}} {\cal D}_H={\cal T}$ where $\cal T$ denotes the depolarizing channel i.e., ${\cal T}(X)=\mathrm{Tr}(X) \frac{\openone}{d}.$
 
%%%%%%%%%%%%%%%%%%%%%%%%%%%%%%%%%%%%% END END END%%%%%%%%%%%%%%%%%%%%%%%%%%%%%%%%%%%%% END END END%%%%%%%%%%%%%%%%%%%%%%%%%%%%%%%%%%%%% END END END%%%%%%%%%%%%%%%%%%%%%%%%%%%%%%%%%%%%% END END END%%%%%%%%%%%%%%%%%%%%%%%%%%%%%%%%%%%%% END END END%%%%%%%%%%%%%%%%%%%%%%%%%%%%%%%%%%%%% END END END%%%%%%%%%%%%%%%%%%%%%%%%%%%%%%%%%%%%% END END END%%%%%%%%%%%%%%%%%%%%%%%%%%%%%%%%%%%%% END END END%%%%%%%%%%%%%%%%%%%%%%%%%%%%%%%%%%%%% END END END%%%%%%%%%%%%%%%%%%%%%%%%%%%%%%%%%%%%% END END END%%%%%%%%%%%%%%%%%%%%%%%%%%%%%%%%%%%%% END END END%%%%%%%%%%%%%%%%%%%%%%%%%%%%%%%%%%%%% END END END%%%%%%%%%%%%%%%%%%%%%%%%%%%%%%%%%%%%% END END END%%%%%%%%%%%%%%%%%%%%%%%%%%%%%%%%%%%%% END END END%%%%%%%%%%%%%%%%%%%%%%%%%%%%%%%%%%%%% END END END%%%%%%%%%%%%%%%%%%%%%%%%%%%%%%%%%%%%% END END END%%%%%%%%%%%%%%%%%%%%%%%%%%%%%%%%%%%%% END END END%%%%%%%%%%%%%%%%%%%%%%%%%%%%%%%%%%%%% END END END%%%%%%%%%%%%%%%%%%%%%%%%%%%%%%%%%%%%% END END END%%%%%%%%%%%%%%%%%%%%%%%%%%%%%%%%%%%%% END END END%%%%%%%%%%%%%%%%%%%%%%%%%%%%%%%%%%%%% END END END%%%%%%%%%%%%%%%%%%%%%%%%%%%%%%%%%%%%% END END END%%%%%%%%%%%%%%%%%%%%%%%%%%%%%%%%%%%%% END END END
%%%%%%%%%%%%%%%%%%%%%%%%%%%%%%%%%%%%% END END END
\end{widetext}
\end{document}